\documentclass[letter,twocolumn]{jpsj3}
\usepackage{txfonts}
\usepackage{bm}

\title{Discontinuous change of viscosity in a sheared granular gas\\
with velocity-dependent restitution}

\author{
    Makoto R. Kikuchi$^1$\thanks{m-kikuchi@st.go.tuat.ac.jp},
    Yuria Kobayashi$^1$, and 
    Satoshi Takada$^1$\thanks{Corresponding author, takada@go.tuat.ac.jp}}
\inst{
    $^1$Department of Mechanical Systems Engineering,
    Tokyo University of Agriculture and Technology,\\
    2-24-16 Naka-cho, Koganei, Tokyo 184-8588, Japan} 

\abst{
We investigate the rheology of a sheared granular gas composed of hard spheres with a velocity-dependent restitution coefficient. 
Using kinetic theory, we derive the shear viscosity and show that it exhibits an S-shaped dependence on the shear rate when the restitution coefficient switches between two values depending on the collision velocity. 
As a result, a discontinuous change of viscosity emerges between low- and high-shear regimes, both characterized by Bagnold-type scaling. 
While the phenomenology resembles the Wyart-Cates scenario for dense suspensions, the present transition arises purely from kinetic effects without frictional contacts or jamming.
}

\begin{document}
\maketitle
Granular flows composed of dissipative particles exhibit rich nonequilibrium behavior far from thermal equilibrium~\cite{Jop06, Boyer11, Midi04, Silbert01, Jenkins85_1, Jenkins85_2, Lutsko05, Saitoh07, Gnoli16}.
Even in dilute gas regimes, inelastic collisions lead to nontrivial transport properties, clustering instabilities, and anomalous energy dissipation~\cite{Jenkins85_1, Jenkins85_2, Brey98}, phenomena that are further enhanced and well documented in moderately dense systems~\cite{Garzo99, Mitarai07, Chialvo13}.
To capture realistic collision processes, models with a velocity-dependent restitution coefficient have been extensively studied~\cite{Brilliantov96, Brilliantov}, motivated by viscoelastic contacts and impact-induced dissipation.

It is worth noting that the discontinuous rheological behavior identified in this study appears when the restitution coefficient in the low-velocity regime is smaller than that in the high-velocity regime. 
Although this condition is opposite to the trend commonly observed in dry granular materials, it may be realized in systems where additional dissipation mechanisms are active at low velocities, such as charged~\cite{Scheffler02, Poschel03} or adhesive particles~\cite{Brilliantov07}, as well as suspension systems where hydrodynamic interactions or lubrication effects modify collision dynamics~\cite{Seto18, ASingh18}. 
Thus, the present setup provides a minimal framework to explore how competing dissipation channels can give rise to nontrivial rheological transitions.

In particular, simplified models in which the restitution coefficient equals almost unity below a threshold collision velocity and takes a constant value above it have been introduced to isolate the effect of impact velocity on energy dissipation, motivated not only by viscoelastic contacts but also by dissipative processes in charged granular gases~\cite{Scheffler02, Poschel03, Takada17, Takada22, Singh18, Singh19}.
For such models, the homogeneous cooling state has been analyzed in detail~\cite{Poschel03, Takada17}, and transport coefficients such as shear viscosity and thermal conductivity have been derived within kinetic theory~\cite{Takada22}.
These studies clarified how velocity-dependent dissipation modifies cooling rates and hydrodynamic coefficients, while remaining within spatially homogeneous or weakly perturbed states.

However, the rheological consequences of velocity-dependent restitution under steady shear have only recently begun to be explored, and the available literature remains very limited~\cite{Yoshii23, Kobayashi25, Kobayashi26}.
Under shear, the collision velocity distribution is strongly modified, suggesting that different dissipation regimes may dominate depending on the shear rate.
This raises the possibility of nontrivial constitutive behavior that cannot be inferred from the homogeneous cooling state alone.

In this Letter, we show that a granular gas with velocity-dependent restitution exhibits an S-shaped constitutive curve and a discontinuous change of viscosity under shear.
Using kinetic theory, we demonstrate that the transition occurs between two Bagnold-type regimes~\cite{Bagnold54} characterized by different effective restitution coefficients.
Although the phenomenology is reminiscent of the Wyart-Cates scenario~\cite{Wyart14, Cates14, Guy19} for dense suspensions, the present mechanism is purely kinetic and does not rely on frictional contacts or jamming.
This suggests that discontinuous rheological responses may arise more generally from the competition between distinct dissipative mechanisms, even in the absence of frictional contacts.

We consider a dilute frictionless granular gas composed of monodisperse hard spheres with mass $m$ and diameter $d$.
The system is sufficiently dilute, and the particle volume fraction is fixed at $\varphi=0.01$.
The position and velocity of the $i$-th particle are denoted by $\bm{r}_i$ and $\bm{v}_i$, respectively.

The normal restitution coefficient is assumed to depend on the normal component of the relative velocity $v_n$ between two colliding particles as
\begin{equation}
    e(v_n)=e_1 - (e_1-e_2)\Theta(v_n-v_\mathrm{c}),
    \label{eq:def_COR}
\end{equation}
where $v_\mathrm{c}$ is the threshold velocity.
When two particles with velocities $(\bm{v}_1, \bm{v}_2)$ collide, their post-collisional velocities are given by
\begin{equation}
    \bm{v}_1^\prime = \bm{v}_1 - \frac{1+e}{2}\left(\bm{v}_{12}\cdot \hat{\bm{k}}\right)\hat{\bm{k}},\quad
    \bm{v}_2^\prime = \bm{v}_2 + \frac{1+e}{2}\left(\bm{v}_{12}\cdot \hat{\bm{k}}\right)\hat{\bm{k}},
\end{equation}
where $\bm{v}_{12}\equiv \bm{v}_1-\bm{v}_2$, and $\hat{\bm{k}}\equiv (\bm{r}_1-\bm{r}_2)/|\bm{r}_1-\bm{r}_2|$ is the unit vector along the line connecting the centers of the two particles at contact.

A uniform shear flow with shear rate $\dot\gamma$ is imposed on the system.
We assume that the system remains spatially uniform in the sheared state, neglecting spatial fluctuations.
Accordingly, the dynamics of each particle is described in terms of the peculiar velocity
\begin{equation}
    \bm{V}_i\equiv \bm{v}_i - \dot\gamma y_i \hat{e}_x,
\end{equation}
where $\hat{\bm{e}}_x$ is the unit vector in the $x$-direction.

In the uniformly sheared state, the velocity distribution function $f(\bm{V},t)$ obeys the Boltzmann equation~\cite{Chapman, Garzo, Santos04}
\begin{equation}
    \left(\frac{\partial}{\partial t}
    -\dot\gamma V_y \frac{\partial}{\partial V_x}\right)f(\bm{V}, t)=J(f,f),
    \label{eq:Boltzmann}
\end{equation}
where $J(f,f)$ is the collision integral accounting for binary collisions.
It is given by
\begin{align}
    J(f,f)
    &= \int \mathrm{d}\bm{V}_2 \int \mathrm{d}\hat{\bm{k}}
    \Theta\left(\bm{V}_{12}\cdot \hat{\bm{k}}\right)
    \left(\bm{V}_{12}\cdot \hat{\bm{k}}\right)\nonumber\\
    &\hspace{1em}\times
    \left[\frac{1}{e^2}f(\bm{V}_1^{\prime\prime},t)f(\bm{V}_2^{\prime\prime},t)
    - f(\bm{V}_1, t)f(\bm{V}_2,t)\right],
\end{align}
where $(\bm{V}_1^{\prime\prime}, \bm{V}_2^{\prime\prime})$ denote the pre-collisional velocities that lead to $(\bm{V}_1, \bm{V}_2)$ after collision.

To analyze the rheology of the granular gas, it is more convenient to consider the time evolution of the kinetic contribution to the stress tensor, defined as the second velocity moment of the distribution function,
\begin{equation}
    P_{\alpha\beta}\equiv \int \mathrm{d}\bm{V}mV_\alpha V_\beta f(\bm{V}, t).
\end{equation}
Although the stress tensor generally contains both kinetic and collisional contributions, the latter can be neglected in the dilute limit considered here.
From Eq.~\eqref{eq:Boltzmann}, one obtains~\cite{Garzo, Hayakawa19, Hayakawa17, Santos04, Kobayashi25, Kobayashi26, Sugimoto20, Takada25}
\begin{equation}
    \frac{\partial P_{\alpha\beta}}{\partial t}
    + \dot\gamma \left(\delta_{\alpha x}P_{y\beta}
    + \delta_{\beta x}P_{y\alpha}\right)
    = -\Lambda_{\alpha\beta},
    \label{eq:P_evol}
\end{equation}
where
\begin{equation}
    \Lambda_{\alpha\beta}
    \equiv -m\int \mathrm{d}\bm{V} V_\alpha V_\beta 
    J(f,f),
    \label{eq:def_Lambda}
\end{equation}
represents the collisional contribution to the stress evolution.

Equation~\eqref{eq:P_evol} formally consists of nine components.
However, symmetry requires $P_{\alpha\beta}=P_{\beta\alpha}$, and in a uniformly sheared state one has $P_{xz}=P_{yz}=0$.
Furthermore, in the dilute limit, $P_{yy}\simeq P_{zz}$ holds~\cite{Santos04, Hayakawa19}.
Therefore, it is sufficient to consider only three independent components.
We define the granular temperature $T$ and the temperature anisotropy $\Delta T$ as
\begin{equation}
    T\equiv \frac{P_{xx}+P_{yy}+P_{zz}}{3n},\quad
    \Delta T\equiv \frac{P_{xx}-P_{yy}}{n},
\end{equation}
where $n=6\varphi/(\pi d^3)$ is the number density.
Using these definitions, Eq.~\eqref{eq:P_evol} can be rewritten as
\begin{subequations}\label{eq:dyn_eqs}
\begin{align}
    \frac{\partial T}{\partial t}
    &= -\frac{2}{3n}\dot\gamma P_{xy}
    - \frac{\Lambda_{\alpha\alpha}}{3n},\\
    \frac{\partial \Delta T}{\partial t}
    &= -\frac{2}{n}\dot\gamma P_{xy}
    - \frac{\Lambda_{xx}-\Lambda_{yy}}{n},\\
    \frac{\partial P_{xy}}{\partial t}
    &= -\dot\gamma n\left(T-\frac{1}{3}\Delta T\right)
    - \Lambda_{xy}.
\end{align}
\end{subequations}
By solving this set of coupled equations, the rheological properties of the system can be determined.

Up to this point, no approximation has been introduced.
However, an exact solution of the velocity distribution function $f(\bm{V}, t)$ is not available at present.
Instead, we adopt Grad's approximation~\cite{Grad49},
\begin{equation}
    f(\bm{V},t)=f_\mathrm{M}(\bm{V},t)
    \left[1+\frac{m}{2T}\left(\frac{P_{\alpha\beta}}{nT}-\delta_{\alpha\beta}\right)V_\alpha V_\beta\right],
    \label{eq:Grad}
\end{equation}
where
\begin{equation}
    f_\mathrm{M}(\bm{V}, t)
    \equiv n\left(\frac{m}{2\pi T}\right)^{3/2}
    \exp\left(-\frac{mV^2}{2T}\right),
\end{equation}
is the Maxwellian distribution.
It is known that analyses based on this approximation reproduce simulation results with high accuracy and without fitting parameters for systems of hard spheres with a constant restitution coefficient~\cite{Santos04, Hayakawa19, Hayakawa17} as well as for systems with harmonic or other interparticle interactions~\cite{Sugimoto20, Takada18}.
We therefore employ the same approximation in the present study.

Using the Grad approximation~\eqref{eq:Grad}, the collisional term defined in Eq.~\eqref{eq:def_Lambda} can be evaluated explicitly as
\begin{equation}
    \Lambda_{\alpha\beta}
    = \zeta nT\delta_{\alpha\beta}
    + \nu \left(P_{\alpha\beta}-nT\delta_{\alpha\beta}\right).
\end{equation}
Here, $\zeta$ and $\nu$ represent the energy dissipation rate and an effective collision frequency, respectively, and are given by
\begin{equation}
    \zeta
    \equiv \frac{\sqrt{2\pi}}{3}nd^2 v_\mathrm{T} \Omega_5^{(1)},\ 
    \nu
    \equiv \frac{\sqrt{2\pi}}{15}nd^2 v_\mathrm{T}
    \left(\Omega_7^{(1)}+\frac{3}{2}\Omega_7^{(2)}\right),
\end{equation}
where $v_\mathrm{T}\equiv \sqrt{2T/m}$.
The quantities $\Omega_n^{(i)}$ ($i=1, 2$) are defined as~\cite{Chapman, Hirschfelder}
\begin{subequations}
\begin{align}
    \Omega_n^{(1)}
    &\equiv \int_0^\infty \mathrm{d}g
    \int_0^1 \mathrm{d}(\cos\theta)
    \left[1-e(g\cos\theta)^2\right]
    g^n \cos^3\theta \mathrm{e}^{-g^2/2},\\
    \Omega_n^{(2)}
    &\equiv \int_0^\infty \mathrm{d}g
    \int_0^1 \mathrm{d}(\cos\theta)\nonumber\\
    &\hspace{1em}\times 
    \left[1+e(g\cos\theta)\right]^2
    g^n \sin^2\theta\cos^2\theta \mathrm{e}^{-g^2/2}.
\end{align}
\end{subequations}
These quantities characterize binary collisions and are determined by the interparticle interaction.
In general, evaluating these parameters for a given interaction potential is essentially equivalent to determining the rheological properties of the system.

For the velocity-dependent restitution coefficient defined in Eq.~\eqref{eq:def_COR}, these integrals can be evaluated analytically, yielding
\begin{subequations}
\begin{align}
    \Omega_5^{(1)}
    &= 2\left[\left(1-e_1^2\right)
    + \left(e_1^2-e_2^2\right)\left(1+x\right)\mathrm{e}^{-x}\right],\\
    \Omega_7^{(1)}
    &= 12\left[\left(1-e_1^2\right)
    + \left(e_1^2-e_2^2\right)
    \left(1+x+\frac{1}{3}x^2\right)\mathrm{e}^{-x}\right],\\
    \Omega_7^{(2)}
    &= 4\left[\left(1+e_1\right)^2
    - \left(e_1-e_2\right)
    \left(2+e_1+e_2\right)
    \left(1+x\right)\mathrm{e}^{-x}\right],
\end{align}
\end{subequations}
where the dimensionless parameter $x$, corresponding to the threshold velocity $v_\mathrm{c}$, is defined as
\begin{equation}
    x \equiv \frac{mv_\mathrm{c}^2}{4T}.
\end{equation}

Focusing on the steady state, the coupled equations~\eqref{eq:dyn_eqs} can be solved to give
\begin{subequations}\label{eq:steady}
\begin{align}
    &\dot\gamma=\nu \sqrt{\frac{3}{2}\frac{\zeta}{\nu-\zeta}},\quad
    \Delta T=\frac{3\zeta}{\nu} T,\\
    &P_{xy}= -\frac{nT}{\nu}\sqrt{\frac{3}{2}\zeta(\nu-\zeta)},
\end{align}
\end{subequations}
from which the shear viscosity $\eta$ is obtained as
\begin{equation}
    \eta=-\frac{P_{xy}}{\dot\gamma}
    = nT \frac{\nu-\zeta}{\nu^2}.
\end{equation}
Thus, all macroscopic quantities in the steady state can be expressed solely as functions of the granular temperature $T$.

\begin{figure}[htbp]
    \centering
    \includegraphics[width=\linewidth]{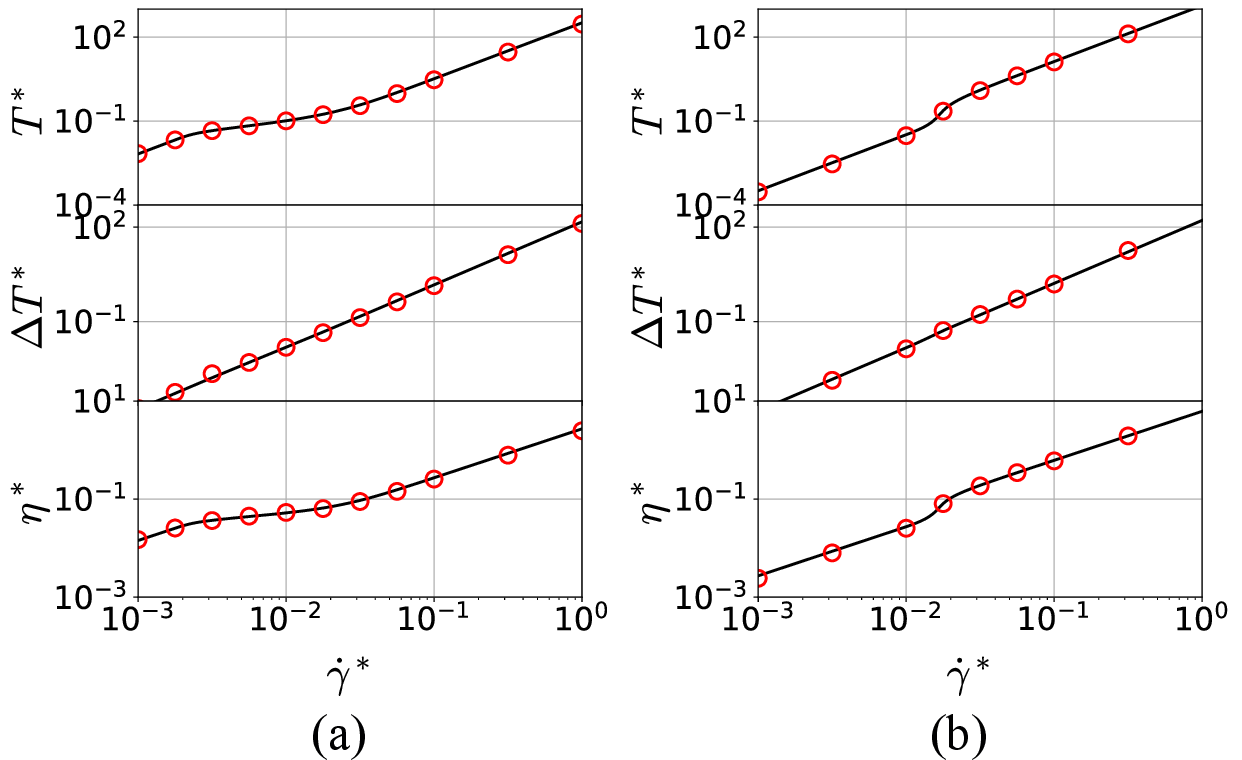}
    \caption{Typical flow curves of the temperature, anisotropic temperature, and viscosity for (a) $e_1>e_2$ ($(e_1, e_2)=(0.99, 0.80)$) and (b) $e_1<e_2$ ($(e_1, e_2)=(0.80, 0.95)$).
    Here, we have introduced the dimensionless quantities $T^*\equiv T/(mv_\mathrm{c}^2)$, $\Delta T^*\equiv \Delta T/(mv_\mathrm{c}^2)$, and $\eta^*\equiv \eta d^2/(mv_\mathrm{c}^2)$.
    Symbols represent results of event-driven molecular dynamics simulations.}
	\label{fig:rheology}
\end{figure}

Figure~\ref{fig:rheology} shows typical shear-rate dependences of the temperature, the temperature anisotropy, and the shear viscosity for the cases $e_1>e_2$ (Fig.~\ref{fig:rheology}(a)) and $e_1<e_2$ (Fig.~\ref{fig:rheology}(b)).
To validate the theoretical predictions, we also performed event-driven molecular dynamics simulations of hard spheres with velocity-dependent restitution.
In the low-shear (high-shear) limit, almost all collisions occur with relative velocities smaller (larger) than the threshold velocity $v_\mathrm{c}$.
In these limits, the system can therefore be regarded as a gas of hard spheres with a constant restitution coefficient equal to $e_1$ ($e_2$), respectively.
Accordingly, the shear-rate dependences of the temperature and related quantities are described by the Bagnold scaling laws~\cite{Bagnold54},
\begin{subequations}\label{eq:Bagnold}
\begin{align}
    T^{(\mathrm{HC})}(e, \dot\gamma)
    &= \frac{5(2+e)}{12\pi(1-e)(1+e)^2(3-e)^2}
    \frac{m}{n^2d^4}
    \dot\gamma^2,\\
    \Delta T^{(\mathrm{HC})}(e, \dot\gamma)
    &= \frac{25(2+e)}{12\pi(1+e)^2(3-e)^3}
    \frac{m}{n^2d^4}
    \dot\gamma^2,\\
    \eta^{(\mathrm{HC})}(e, \dot\gamma)
    &= \frac{5\sqrt{15}(2+e)^{3/2}}{36\pi(1-e)^{1/2}(1+e)^2(3-e)^3}
    \frac{m}{nd^4}\dot\gamma.
\end{align}
\end{subequations}
Indeed, the theoretical curves in Fig.~\ref{fig:rheology} asymptotically approach these Bagnold scalings in both the low- and high-shear regimes, confirming the above interpretation.
In contrast, in the intermediate shear regime, where a significant fraction of particles collide with velocities around $v_\mathrm{c}$, both characteristics associated with $e_1$ and $e_2$ coexist.
As a consequence, the two asymptotic branches are smoothly connected through a bent crossover region bridging the low- and high-shear limits.

\begin{figure}
    \centering
    \includegraphics[width=\linewidth]{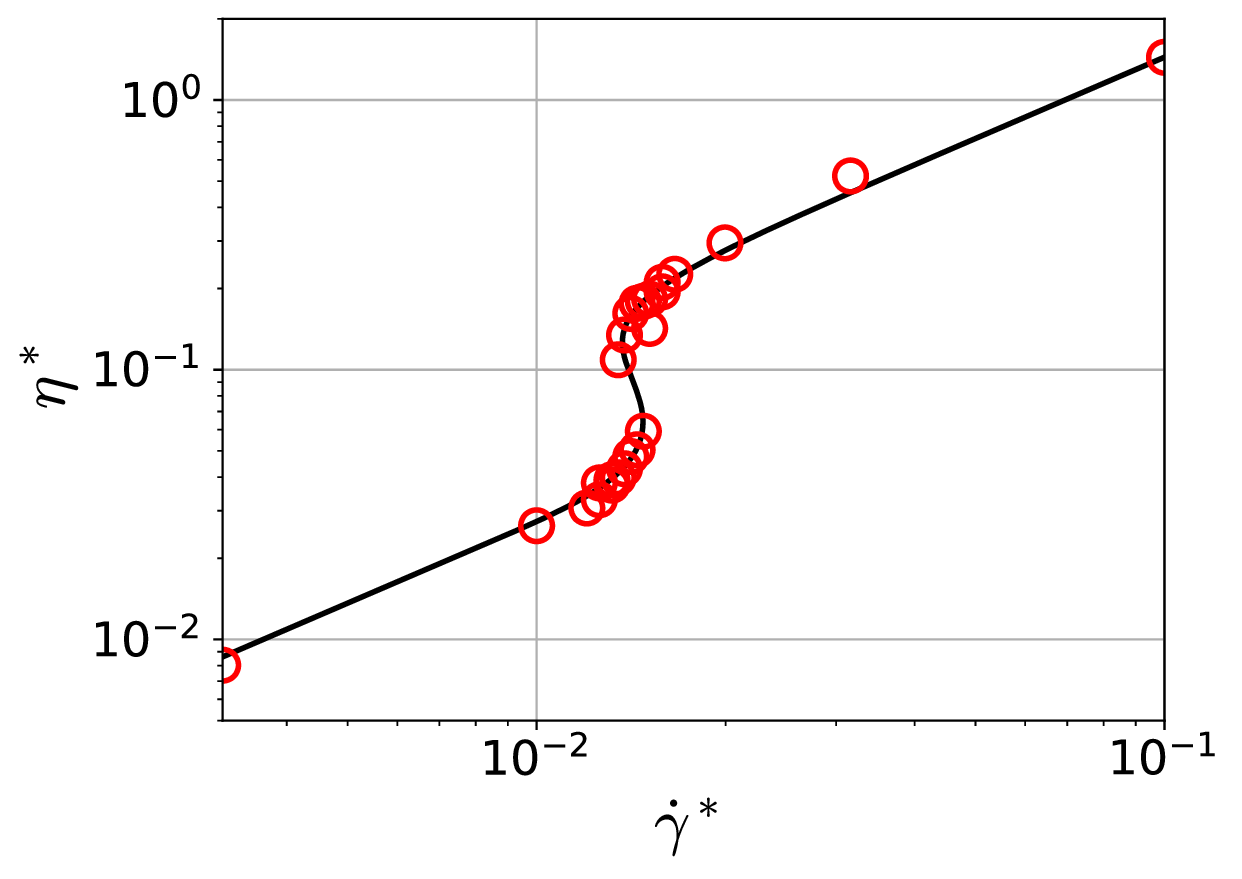}
    \caption{Shear-rate dependence of the viscosity exhibiting an S-shaped curve for $(e_1, e_2)=(0.80, 0.99)$.
    Symbols represent results of event-driven molecular dynamics simulations.}
	\label{fig:S_shape}
\end{figure}

A particularly intriguing feature emerges for the case $e_1<e_2$.
For suitable choices of $(e_1, e_2)$, the shear-rate dependence of the viscosity exhibits an S-shaped curve, as shown in Fig.~\ref{fig:S_shape}.
This implies that, within a certain range of shear rates, three distinct values of the viscosity coexist for a single imposed shear rate.
This behavior closely resembles discontinuous shear thickening observed in suspension systems.
For instance, when the shear rate is gradually increased along the lower branch, the slope of the viscosity diverges at a critical point, beyond which no stable state exists and the system jumps discontinuously to the upper branch.

Despite this phenomenological similarity, the underlying physics is fundamentally different.
In dense suspensions, the S-shaped flow curve arises from the coexistence of a fluid-like flowing branch and a jammed, solid-like branch.
This mechanism has been successfully captured by the Wyart-Cates model, in which frictionless and frictional contact states form two distinct branches, and the fraction of frictional contacts controls the nonlinear divergence of the viscosity near jamming.
In contrast, the present system consists exclusively of a dilute granular gas in both branches, with no distinction other than the effective restitution coefficient governing collisions.
In this sense, although the resulting rheology is reminiscent of a Wyart-Cates-like scenario~\cite{Wyart14, Cates14, Guy19}, it represents a qualitatively new type of discontinuous transition driven purely by velocity-dependent dissipation.

Finally, we note that such S-shaped behavior does not appear for $e_1>e_2$, and in particular no shear thinning is observed in this case.
This can be understood straightforwardly from the temperature--shear-rate relation.
Equation~\eqref{eq:steady} expresses the shear rate as a single-valued function of the temperature.
If shear thinning were present, multiple shear rates would correspond to a single temperature, which contradicts this functional dependence.
Therefore, although the slope of the flow curve may approach zero, it never becomes negative.

\begin{figure}[htbp]
    \centering
    \includegraphics[width=\linewidth]{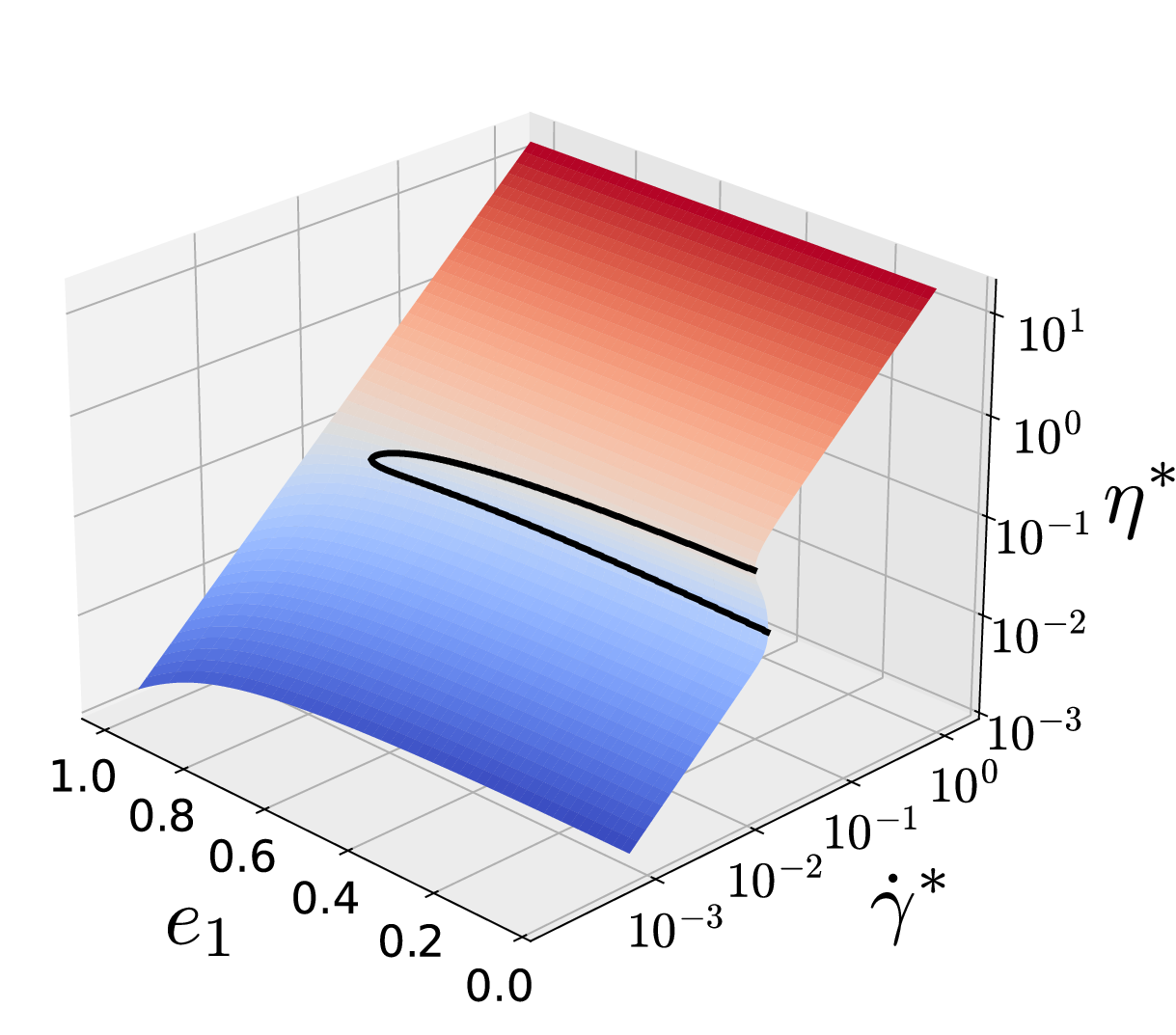}
    \caption{Three-dimensional plot of the shear-rate dependence of the viscosity for fixed $e_2=0.99$ and varying $e_1$.
    The solid line connects the points at which $\partial \eta^*/\partial \dot\gamma^*$ diverges.}
	\label{fig:3D}
\end{figure}

We now investigate under what conditions the S-shaped constitutive relation emerges in the case $e_1<e_2$.
Figure~\ref{fig:3D} presents a three-dimensional plot of the viscosity $\eta$ as a function of the shear rate $\dot\gamma$, while continuously varying $e_1$ at fixed $e_2$.
For sufficiently small values of $e_1$, the curve $\eta(\dot\gamma)$ exhibits a pronounced S-shaped structure, implying the existence of a shear-rate interval where the viscosity becomes triple-valued.
As $e_1$ increases, however, this multivalued region gradually shrinks, and eventually the S-shaped structure disappears, leaving the viscosity as a single-valued function of the shear rate.

The solid line in Fig.~\ref{fig:3D} connects the points where the slope $\partial\eta/\partial \dot\gamma$ diverges, and thus represents the boundary of the parameter region in which the S-shaped structure can exist~\cite{Hayakawa17}.
This result indicates that the competition between two effective dissipative states-associated with low- and high-collision, velocity regimes, weakens as $e_1$ approaches $e_2$, thereby suppressing the discontinuous rheological response.

\begin{figure}[htbp]
    \centering
    \includegraphics[width=\linewidth]{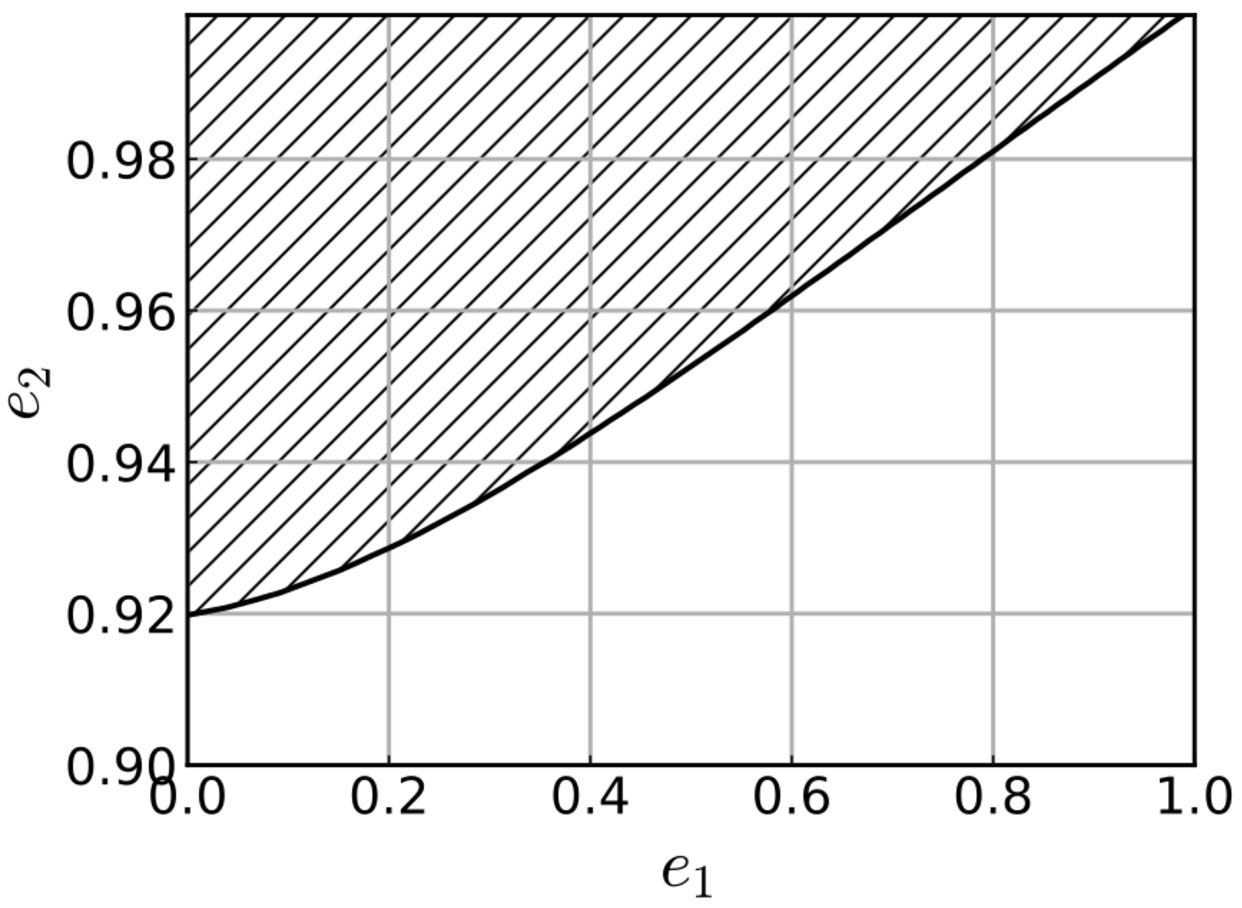}
    \caption{Phase diagram in the $(e_1,e_2)$ plane indicating whether discontinuous shear thickening occurs.
    The shaded region corresponds to the existence of an S-shaped constitutive relation.}
	\label{fig:threshold}
\end{figure}

Figure~\ref{fig:threshold} shows a phase diagram in the $(e_1, e_2)$ plane, indicating whether an S-shaped constitutive relation, namely, a multivalued viscosity, appears.
In the shaded region, $\eta(\dot\gamma)$ remains single-valued for all shear rates, and no S-shaped structure is observed.
In contrast, in the unshaded region, the viscosity becomes multivalued within an appropriate shear-rate range, allowing for a discontinuous transition.
As clearly seen in Fig.~\ref{fig:threshold}, the S-shaped structure appears only under the condition $e_1<e_2$, and only when the difference between $e_1$ and $e_2$ is sufficiently large.
This demonstrates that a pronounced contrast between the effective dissipation strengths in the low- and high-collision-velocity regimes is essential for the competition between the two Bagnold-type states to become manifest.

In summary, we have performed a kinetic-theoretical analysis of the shear rheology of a dilute granular gas with a velocity-dependent coefficient of restitution, and demonstrated the emergence of an S-shaped constitutive relation and a discontinuous change in the viscosity.
This transition originates from the competition between two Bagnold-type states characterized by different restitution coefficients, and is obtained without invoking frictional contacts or jamming, relying only on minimal microscopic ingredients.

Although the present analysis is restricted to the dilute limit, it is an intriguing future problem to investigate whether the discontinuous transition persists in denser systems by incorporating density effects. 
In such regimes, additional contributions such as collisional transfer, enduring contacts, and spatial correlations may play important roles and modify the rheological response. 
We also note that the present model neglects rotational degrees of freedom and assumes a simplified stepwise restitution law, and it would be interesting to examine the robustness of the transition against more realistic descriptions.

Furthermore, since the present mechanism is based on competing dissipation processes depending on collision velocity, similar behavior is expected to arise in systems with more realistic velocity-dependent restitution models.

\begin{acknowledgment}
This work is partially supported by the Grant-in-Aid of MEXT for Scientific Research (Grant No.~JP24K06974, No.~JP24K07193, and No.~JP25K01063).
\end{acknowledgment}


\end{document}